\documentclass[aps,prb,twocolumn,amsmath,amssymb,superscriptaddress,floatfix]{revtex4-1}

\usepackage{graphicx}
\usepackage{multirow}
\usepackage{xcolor}
\usepackage{bm}
\usepackage{amsfonts,amssymb,amsmath}
\usepackage{graphicx,dcolumn,bm,color,braket,slashed}
\usepackage{times} 
\usepackage{comment}
\usepackage{array}
\usepackage{siunitx}
\def\ket#1{|#1\rangle }
\def\bra#1{\langle #1 |}

\newcommand\abs[1]{\left|#1\right|}

\begin{document}

\title{Low-energy electrodynamics of Dirac semimetal phases in the doped Mott insulator Sr$_2$IrO$_4$}

\author{Sun-Woo Kim}
\affiliation{Department of Physics and Research Institute for Natural Science, Hanyang University, Seoul 04763, Korea}

\author{Myungjun Kang}
\affiliation{Department of Physics and Research Institute for Natural Science, Hanyang University, Seoul 04763, Korea}

\author{Sangmo Cheon}
\email{sangmocheon@hanyang.ac.kr}
\affiliation{Department of Physics and Research Institute for Natural Science, Hanyang University, Seoul 04763, Korea}

\date{\today}
\begin{abstract}
Correlated Dirac semimetal phases emerge in lightly doped (Tb- or La-doped) Mott insulator Sr$_2$IrO$_4$, where a $d$-wave symmetry-breaking order underlying a pseudogap plays a crucial role in determining the nature of Dirac degeneracy, i.e., whether it is a Dirac line node or Dirac point node.
Here, using a realistic five-orbital tight-binding model with a Hubbard $U$ and a semiclassical Boltzmann transport theory, we systematically study the low-energy electrodynamic properties of the Dirac semimetal phases in the paramagnetic lightly doped  Sr$_2$IrO$_4$.
We investigate the effects of the $d$-wave electronic order and electron doping concentration on the electronic band structures and optical properties of various Dirac semimetal phases.
We calculate the intraband optical conductivity and obtain electrodynamic parameters of dc conductivity, scattering rate, and Drude weight for three Dirac semimetal phases: two are Dirac point-node states observed in the 3\% Tb-doped and 5\% La-doped Sr$_2$IrO$_4$, and the other is a Dirac line-node state.
Our results show that the temperature dependence of the electrodynamic parameters is strong in the Tb-doped system while weak in the La-doped and Dirac line-node systems, which are consistent with available experimental data.
Moreover, using the low-energy effective theory, we also compare the temperature-dependent screening effect in the Tb- and La-doped systems using graphene as a reference.
Our work provides valuable insight for understanding the transport and optical properties of correlated Dirac semimetal phases in the doped Sr$_2$IrO$_4$. 
\end{abstract}

\maketitle

\newpage

\section{Introduction}

Since the discovery of graphene~\cite{castro_neto_electronic_2009}, there has been a great deal of work on the transport and optical properties of graphene as well as various Dirac semimetals due to the unique massless and chiral nature of Dirac fermions~\cite{das_sarma_electronic_2011,armitage_weyl_2018}.
For instance, high electron mobility was observed in graphene~\cite{bolotin_ultrahigh_2008} and the correlated Dirac semimetal CaIrO$_3$~\cite{fujioka_strong-correlation_2019}. 
Moreover, two distinct behaviors of resistivity with respect to temperature (i.e., metallic or insulating) depending on the carrier density~\cite{bolotin_temperature-dependent_2008} and even nonmonotonic temperature dependence of resistivity~\cite{heo_nonmonotonic_2011} in graphene were reported.
These were attributed to the competition of different scattering mechanisms (e.g., charged impurities, short-range impurities, and phonons)~\cite{hwang_screening-induced_2009,li_disorder-induced_2011,das_sarma_density-dependent_2013}.
The unusual temperature dependence of resistivity was also observed in other Dirac semimetals~\cite{li_giant_2015,xiong_evidence_2015,li_chiral_2016}.
For low-energy optical properties, the free-carrier electrodynamics showing Drude response with small scattering rate was observed for various Dirac semimetals~\cite{chen_optical_2015,xu_optical_2016,park_electrodynamic_2017,crassee_nonuniform_2018}. 

Recently, the emergence of a correlated Dirac semimetal is demonstrated in the lightly doped Mott insulators 3\% Tb- and 5\% La-doped Sr$_2$IrO$_4$~\cite{de_la_torre_collapse_2015,zhou_correlation_2017,han_nonsymmorphic_2020}.
Interestingly, it is revealed that the nature of Dirac degeneracy protected by the nonsymmorphic symmetry depends on the correlation-induced $d$-wave symmetry-breaking order underlying the pseudogap~\cite{zhou_correlation_2017,han_nonsymmorphic_2020}.
The $d$-wave electronic order $\Delta_d$ plays an important role as an order parameter for determining the territory of Dirac degeneracy, i.e., whether it is a Dirac line node along the whole Brillouin zone (BZ) boundary for $\Delta_d=0$ or Dirac point nodes at the $X$ and $Y$ points for $\Delta_d\neq0$ [see Figs. 1(a) and 1(b)]. 
The different natures of Dirac quasiparticles are expected to give rise to distinct electrodynamic properties, leading to the observation of the Dirac point-node state in the 3\% Tb-doped Sr$_2$IrO$_4$ by terahertz experiments~\cite{han_nonsymmorphic_2020}.
However, the detailed role of the $d$-wave order parameter and carrier doping on the low-energy electrodynamics of Dirac semimetal phases in doped Sr$_2$IrO$_4$ is not understood yet and thus a systematic study for Dirac semimetal phases in various doped Sr$_2$IrO$_4$ is required.

In this work, we theoretically investigate the low-energy electrodynamic properties of various Dirac semimetal phases in paramagnetic lightly doped Sr$_2$IrO$_4$.
Using a realistic five-orbital tight-binding model derived from density-functional theory calculations and a mean-field Hubbard model, we study the effects of the $d$-wave order parameter and electron doping on the electronic structures in the paramagnetic doped Sr$_2$IrO$_4$.
Then, by using the semiclassical Boltzmann transport theory and performing both numerical and analytical calculations, we present the temperature-dependent dc conductivity, scattering rate, and Drude weight, especially for a Dirac line-node state in the undoped system and Dirac point-node states in the 3\% Tb-doped and 5\% La-doped systems.
We also compare our results with available dc and terahertz experiments.
Finally, by constructing the low-energy effective Hamiltonian, we investigate the temperature-dependent screening effect in the Tb- and La-doped systems and compare it to that studied in graphene~\cite{hwang_screening-induced_2009,das_sarma_electronic_2011}.

\section{Methods}
\subsection{Model Hamiltonian}

Upon Tb or La doping, the
antiferromagnetic Mott insulating state of the parent Sr$_2$IrO$_4$~\cite{kim_novel_2008,kim_phase-sensitive_2009} becomes a paramagnetic metallic state with a definite Fermi surface even at low doping concentration, as observed in the 3\% Tb-doped~\cite{wang_decoupling_2015,han_nonsymmorphic_2020} and 5\% La-doped~\cite{de_la_torre_collapse_2015} systems, Sr$_2$(Ir$_{0.97}$Tb$_{0.03}$)O$_4$ and (Sr$_{0.95}$La$_{0.05}$)$_2$IrO$_4$.
Moreover, since the layered Sr$_2$IrO$_4$ system has two-dimensional (2D) electronic structures mainly arising from the Ir atom ~\cite{wang_twisted_2011,carter_theory_2013}, we focus on the 2D paramagnetic metallic single-layer Sr$_2$IrO$_4$ throughout this work.
We adopt the 2D five-orbital tight-binding (TB) model including the spin-orbit coupling (SOC),
TB+SOC Hamiltonian $H_0$~\cite{zhou_correlation_2017}, which is given by
\begin{align}
\begin{split}
H_0= \sum_{ij,\mu\nu,\sigma} t_{ij}^{\mu\nu,\sigma} c_{i\mu\sigma}^{\dagger} c_{j\nu\sigma} 
+ \sum_{i,\mu,\sigma} \epsilon_{\mu} c_{i\mu\sigma}^{\dagger} c_{i\mu\sigma} 
\\ + \sum_{i,\mu\nu,\sigma\sigma'} \lambda_{SOC} \bra{\mu}\mathbf{L}\ket{\nu} \cdot \bra{\sigma}\mathbf{S}\ket{\sigma'} c_{i\mu\sigma}^{\dagger} c_{i\nu\sigma'},
\end{split}
\end{align}
where $c_{i\mu\sigma}^{\dagger}$ is the creation operator for an electron with spin $\sigma$ in the $\mu$th orbital at site $i$,
$\lambda_{SOC}$ is a SOC parameter, 
and $t_{ij}^{\mu\nu,\sigma}$ is the complex hopping integrals between sites $i$ and $j$ of up to the fifth nearest neighbors 
as given in Ref. \cite{zhou_correlation_2017}, which was derived from density-functional theory calculations.
The crystalline electric field effects are taken into account in the on-site energy term $\epsilon_{\mu =( d_{yz}, d_{zx}, d_{xy}, d_{3z^2-r^2}, d_{x^2-y^2})} = (0,0,202,3054,3831)$ meV.
Here, $\mathbf{L}$ and $\mathbf{S}$ are the orbital and spin angular momentum operators, respectively.

For the electron correlation effect, we consider the five-orbital Hubbard Hamiltonian $H_U$ given by, 
\begin{align}
\begin{split}
H_U  = U\sum_{i,\mu} \hat{n}_{i\mu\uparrow} \hat{n}_{i\mu\downarrow}
+ (U'-J/2)\sum_{i,\mu<\nu} \hat{n}_{i\mu} \hat{n}_{i\nu}
 \\
-J \sum_{i,\mu\neq\nu} \mathbf{S}_{i\mu} \cdot \mathbf{S}_{i\nu}
+ J \sum_{i,\mu\neq\nu} c_{i\mu\uparrow}^{\dagger} c_{i\mu\downarrow}^{\dagger} c_{i\nu\downarrow} c_{i\nu\uparrow},
\end{split}
\end{align}
where $\hat{n}$ is the density operator, $U$ and $U'$ are the local intraorbital and interorbital Coulomb repulsions, respectively, and $J$ is the Hund's rule coupling with $U=U'+2J$. 
We use local intraorbital Coulomb repulsion $U=1.4$ and $1.6$ eV on the Ir site for the 5\% La-doped (electron filling at Ir site $f=5.1$) and 3\% Tb-doped ($f=5.01$) systems, respectively, which is a calculated value using constrained random-phase approximation (cRPA) \cite{liu_electron_2016}.
As pointed out for the parent Sr$_2$IrO$_4$ compound \cite{zhou_correlation_2017}, 
we note that the renormalization of SOC due to $U$ is also significant in the 5\% La-doped and 3\% Tb-doped systems as $\lambda_{\text{SOC}}^{\text{eff}} =732$ and $786$ meV, respectively. 
For Hund’s rule coupling $J$, it is known that $J$ favors high-spin states and affects the magnetic anisotropy in the antiferromagnetic state in Sr$_2$IrO$_4$~\cite{jackeli_mott_2009,watanabe_microscopic_2010,zhou_electron_2018}. Since we do not consider such effects in this work, we set $J=0$ to ignore Hund’s rule coupling.

To account for the observed pseudogap~\cite{de_la_torre_collapse_2015,han_nonsymmorphic_2020}, we introduce the $d$-wave spin-orbit density wave ($d$-SODW) order~\cite{zhou_correlation_2017} as a symmetry-breaking order compatible with the symmetry of our doped systems.
It was suggested that the $d$-SODW order has an electronic origin arising from the intersite Coulomb interaction \cite{zhou_correlation_2017}.
The $d$-SODW Hamiltonian $H_\Delta$ in the $J_{\text{eff}}=1/2$ basis is given by 
\begin{align}
\begin{split}
H_\Delta= i\Delta_d \sum_{i\in \text{Ir}_\text{A},\sigma = \pm} \sum_{j=i+\delta} (-1)^{i_y+j_y} \sigma \gamma_{i,\sigma}^{\dagger} \gamma_{j,\sigma}
+ \text{H.c.},
\end{split}
\end{align}
where $\Delta_d$ is a $d$-SODW order parameter, $\delta=\pm\hat{x},\pm\hat{y}$, $(-1)^{i_y+j_y}$ is the nearest-neighbor $d$-wave form factor, and $\gamma_{\pm}^{\dagger}\ket{0}=\ket{J_{\text{eff}}=1/2,J_z=\pm1/2}=\frac{1}{\sqrt{3}}(i\ket{d_{xy},\pm} \pm i \ket{d_{yz},\mp}-\ket{d_{zx},\mp})$. 

The total Hamiltonian $H$ is then given by
\begin{align}
\begin{split}
H  = H_0+H_U+H_\Delta, 
\end{split}
\end{align}
which is solved using the mean-field approximation in this work.
From now on, we refer to it as the TB+SOC+$U$+$\Delta$ method.
Note that the TB+SOC+$U$+$\Delta$ method was originally applied to the La-doped system~\cite{zhou_correlation_2017} where
it could successfully explain the angle resolved photoemission spectroscopy (ARPES) data~\cite{de_la_torre_collapse_2015}. 
For the Tb-doped system, although the current model does not incorporate the Tb $4f$ electrons, 
it is confirmed that in the temperature range of our interest ($T\ge50$ K), the doping concentration of 3\% is low enough: We can treat Tb dopants as a perturbation to the system captured by adjusting parameters in the TB+SOC+$U$+$\Delta$ method.
The validity of this method is justified by comparing the theoretical band structures and optical properties with ARPES and terahertz experiments~\cite{han_nonsymmorphic_2020}.

\subsection{Boltzmann transport theory}
To calculate the intraband optical conductivity of Dirac semimetal phases, we use the semiclassical Boltzmann transport theory.
The linearized time-independent Boltzmann transport equation for the distribution function $f({\mathbf k}) = f^{(0)}(\mathbf k) + \delta f ({\mathbf k})$ is given by
\begin{align}
\begin{split}
\label{eq:Boltzmann_Eq_01}
(-e) \mathbf E \cdot \mathbf {v}_{\mathbf k} S^{(0)}(\epsilon)
= \int \frac{d^2 k\rq{}}{ (2\pi)^2} W_{\mathbf{k}\mathbf{k}\rq{}} [\delta f(\mathbf k)-\delta f(\mathbf k\rq{}) ],
\end{split}
\end{align}
where $f^{(0)}(\mathbf k) = \{ \exp [\beta( \epsilon_{\mathbf k} - \mu)] +1 \}^{-1}$ is the Fermi-Dirac distribution at equilibrium with $\beta=1/k_BT$, the energy $\epsilon_{\mathbf{k}}$, and the chemical potential $\mu$;
$\delta f ({\mathbf k}) $ is the deviation proportional to an applied electric field $\mathbf E$; $\mathbf {v}_{\mathbf k}$ is the velocity of an electron; 
$S^{(0)}(\epsilon) = - \frac{ \partial f^{(0)}(\epsilon)}{\partial \epsilon}$;
and $W_{\mathbf{k}\mathbf{k}'}$ is the quantum mechanical transition probability from $\mathbf{k}'$ to $\mathbf{k}$.
Here, $W_{\mathbf{k}\mathbf{k}'}$ is given by Fermi's golden rule:
\begin{align}
\begin{split}
\label{Fermi_Golden}
W_{\mathbf{k}\mathbf{k}'}
= \frac{2\pi}{\hbar} n_{\text{imp}} |V_{\mathbf{k}\mathbf{k}'}|^2 \delta(\epsilon_{\mathbf{k}} - \epsilon_{\mathbf{k}'}),
\end{split}
\end{align}
where $n_{\text{imp}}$ is the impurity density and
$V_{ \mathbf{k}\mathbf{k}' } = \bra{\mathbf{k}} V  \ket{\mathbf{k}'}$ is the matrix element of the scattering potential $V$.
For isotropic single band systems, the relaxation time $\tau_{\mathbf k}$ within the relaxation-time approximation is given by
\begin{align}
\begin{split}
\label{eq:relaxation_time_integral_iso}
\frac{1}{\tau_{\mathbf k}} 
=
\int \frac{d^2 k\rq{}}{ (2\pi)^2}
W_{ \mathbf{k}\mathbf{k}\rq{} }
\left (1 -  \cos \theta_{\mathbf k \mathbf k\rq{}} \right ),
\end{split}
\end{align}
where 
$
\theta_{\mathbf k \mathbf k\rq{}}
$
is the scattering angle between $\mathbf k$ and $\mathbf k\rq{}$.

Since Dirac dispersions in Dirac semimetal phases of our multiband systems are anisotropic, the relaxation time depends on the momentum direction and hence Eq. (\ref{eq:relaxation_time_integral_iso}) for isotropic single band systems needs to be modified.
Up to a linear order of the external electric field $\mathbf E$, we use the following ansatz~\cite{schliemann_anisotropic_2003,vyborny_semiclassical_2009,park_semiclassical_2017,park_semiclassical_2019} for the deviation of the distribution function from the local equilibrium for the $\alpha$th band $\delta f_{{\alpha}} ({\mathbf k})$:
\begin{align}
\begin{split}
\label{eq:ansatz_02}
\delta f_{\alpha} ({\mathbf k}) 
= (-e) \left(   \sum_{i=1}^{2}  \tau^{(i)}_{\mathbf{k}{\alpha}} {v}_{\mathbf {k}{\alpha}}^{(i)}  E^{(i)}    \right ) 
S^{(0)}(\epsilon),
\end{split}
\end{align}
where ${\tau}_{\mathbf{k}{\alpha}}^{(i)},{v}_{\mathbf{k}{\alpha}}^{(i)}, $ and $ E^{(i)}~(i=x,y)$ 
are the $i$th component of the relaxation time, the velocity, and the electric field, respectively.
Then, by inserting Eq. (\ref{eq:ansatz_02}) into Eq. (\ref{eq:Boltzmann_Eq_01}) and matching each coefficient of $ E^{(i)} $, we get the following integral equation for the relaxation time:
\begin{align}
\begin{split}
\label{eq:relaxation_anisotropic}
1 = {\sum_{\alpha\rq{}}} \int \frac{d^2 k\rq{}}{ (2\pi)^2} W_{\mathbf{k}\mathbf{k}\rq{}}^{{\alpha \alpha\rq{}}}
\left(   \tau_{\mathbf{k}{\alpha}}^{(i)} - \frac{v_{\mathbf k\rq{} {\alpha}\rq{}}^{(i)}}{v_{\mathbf{k}{\alpha}}^{(i)}}  \tau_{\mathbf{k}\rq{}{\alpha}\rq{}}^{(i)}  \right ).
\end{split}
\end{align}
This coupled integral equation can be solved by the numerical method~\cite{liu_mobility_2016,park_semiclassical_2017,park_semiclassical_2019}.
In this work, we ignore the intervalley scattering so that $\alpha=\alpha\rq{}$ for deriving the energy dependence of the scattering rate (see Appendix~\ref{Appendix:sec2}) to calculate the optical conductivity given below.

Under an applied time-dependent electric field $ E^{(i)}(t)$, the linearized time-dependent Boltzmann equation along the $i$th direction for the $\alpha$th band within the relaxation-time approximation is given by
\begin{align}
\begin{split}
 e {v}_{\mathbf{k}{\alpha}}^{(i)}  E^{(i)}(t) S^{(0)}(\epsilon) + \frac{ \partial \delta f_{{\alpha}}^{(i)}(\mathbf k, t)}{ \partial t} 
=
- \frac{ \delta f_{{\alpha}}^{(i)}(\mathbf k, t)}{ \tau^{(i)}_{\mathbf{k}{\alpha}}}.
\end{split}
\end{align}
Using the Fourier transformations, we find a solution for $\delta f_{{\alpha}}$ along the $i$th direction as
\begin{align}
\begin{split}
\delta f_{{\alpha}}^{(i)}(\mathbf k, \omega) = 
\frac{(-e) \tau^{(i)}_{\mathbf{k}{\alpha}}  {v}_{\mathbf{k}{\alpha}}^{(i)}E^{(i)}(\omega) }{ 1 - i \omega \tau^{(i)}_{\mathbf{k}{\alpha}}}
S^{(0)}(\epsilon).
\end{split}
\end{align}
The current density along the $i$th direction $J^{(i)}(\omega)$ is then given by
\begin{align}
\begin{split}
J^{(i)}(\omega)  
& = 
g (-e) {\sum_{\alpha}}
\int \frac{d^2 k}{(2 \pi)^2} 
{v}_{\mathbf{k}{\alpha}}^{(i)} \delta f_{{\alpha}}(\mathbf k, \omega)
\\
&\equiv \sum_{j=1}^{2} \sigma_{ij}(\omega)E^{(j)}(\omega)
,
\end{split}
\end{align}
where $g$ is the degeneracy factor ($g=4$ in our systems) and 
$\sigma_{ij}(\omega)$ is the optical conductivity given by 
\begin{align}
\begin{split}
\label{eq:optical_cond}
\sigma_{ij} (\omega)  = 
g e^2 {\sum_{\alpha}}
\int \frac{d^2 k}{(2 \pi)^2} 
\frac{ {v}_{\mathbf{k}{\alpha}}^{(i)} {v}_{\mathbf{k}{\alpha}}^{(j)} }{ \gamma^{(j)}_{\mathbf{k}{\alpha}}- i \omega }
S^{(0)}(\epsilon),
\end{split}
\end{align}
where $ \gamma^{(i)}_{\mathbf{k}{\alpha}} = 1/\tau^{(i)}_{\mathbf{k}{\alpha}} $ is the momentum-dependent scattering rate along the $i$th direction for the $\alpha$th band.
Note that the dc conductivity is $\sigma_{\text{dc}}^{ij}=\sigma_{ij}(\omega=0)$.

\section{Results and discussion}

\subsection{Band structure}

\begin{figure}[ht]
\includegraphics[width=86mm]{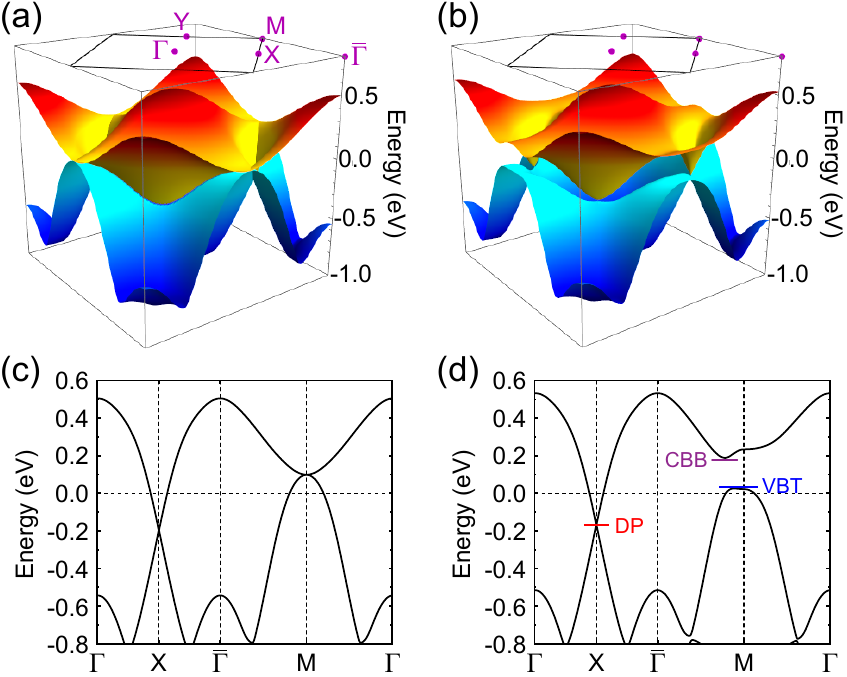} 
\caption{
(a),(b) Three-dimensional (3D) plot of the paramagnetic band structures of two-dimensional (2D) single-layer undoped Sr$_2$IrO$_4$ for the $d$-SODW order parameter (a) $\Delta_d=0$ and (b) $\Delta_d=30$ meV.
Here, only the two bands near the Fermi level are displayed and the BZ is drawn on the top surface.
(c),(d) 2D plot of band structures corresponding to the 3D plot of band structures in (a) and (b), which are plotted along the high-symmetry points indicated in (a). 
The band structures are calculated by using the TB+SOC+$U$+$\Delta$ method with parameters $U=1.6$ and $\lambda_{\text{SOC}}^{\text{eff}} = 786$ meV.
In (d), DP, VBT, and CBB represent a Dirac point, a valence-band top, and a conduction-band bottom, respectively.
}
\end{figure}

Figure~1 shows the representative 3D and 2D plots of band structures of the 2D paramagnetic single-layer Sr$_2$IrO$_4$ obtained by the TB+SOC+$U$+$\Delta$ method.
When the $d$-SODW order parameter $\Delta_d$ is zero, the paramagnetic Sr$_2$IrO$_4$ is a nonsymmorphic Dirac line-node semimetal~\cite{park_two-dimensional_2019,han_nonsymmorphic_2020} where the dispersive fourfold degenerate Dirac nodal line exists along the whole BZ boundary [Fig.~1(a)].
On the other hand, when $\Delta_d\neq0$, a nonsymmorphic Dirac point-node semimetal emerges in the paramagnetic Sr$_2$IrO$_4$ with anisotropic Dirac points at $X$ and $Y$ points [Fig.~1(b)]. 
This nonsymmorphic Dirac point-node semimetal was firstly observed in the 5\% La-doped system~\cite{de_la_torre_collapse_2015,zhou_correlation_2017} and further identified in the 3\% Tb-doped system~\cite{han_nonsymmorphic_2020}. 

In the 2D band structures [Figs.~1(c) and~1(d)], both Dirac semimetals show a linear Dirac dispersion along the $\Gamma-$X$-\bar{\Gamma}$ direction.
At the $M$ point, the fourfold degeneracy is lifted due to the $d$-SODW order. 
The calculated band structures of the two Dirac semimetals are consistent with the previous results~\cite{zhou_correlation_2017,han_nonsymmorphic_2020}.
%
The difference between the two band structures leads to an interesting distinction in electrodynamic properties between the two Dirac semimetals.
For later purposes, we notice the energy positions of a Dirac point (DP), a valence-band top (VBT), and a conduction-band bottom (CBB) [Fig.~1(d)] which rely on the order parameter $\Delta_d$ and electron filling $f$ at the Ir site.
They play significant roles in determining temperature- and frequency-dependent conductivity, as we shall see.

Figure~2 shows the order-parameter dependence of the energy positions of the DP, VBT, and CBB at various fillings.
The filling factor is $f=5+x$, where $x$ denotes the amount of additional electrons induced by doping.
At $\Delta_d=0$, the energy positions of the three points rigidly shift downwards as $x$ increases, because the Fermi level increases with $x$.
As $\Delta_d$ increases, the positions of the VBT (CBB) are monotonically decreasing (increasing), which is a natural consequence of the gap opening at the $M$ point due to the $d$-SODW order.
For the DP positions at all four values of $x$, they saturate at the critical magnitudes of the order parameter ($\Delta_d^{c} \sim 90$ meV), where the position of the VBT becomes lower than that of the DP.
The saturated values of the DP are decided by the fillings $f$: When we integrate the density of states  from the DP to the Fermi level for $\Delta_d \geqq \Delta_d^{c}$, it yields the exact amount of extra electrons $x$. 
Note that, interestingly, in certain conditions ($x=0$ and $\Delta_d\geqq90$ meV), the Dirac point can be located at the Fermi level like pristine graphene.

\begin{figure}[t]
\includegraphics[width=86mm]{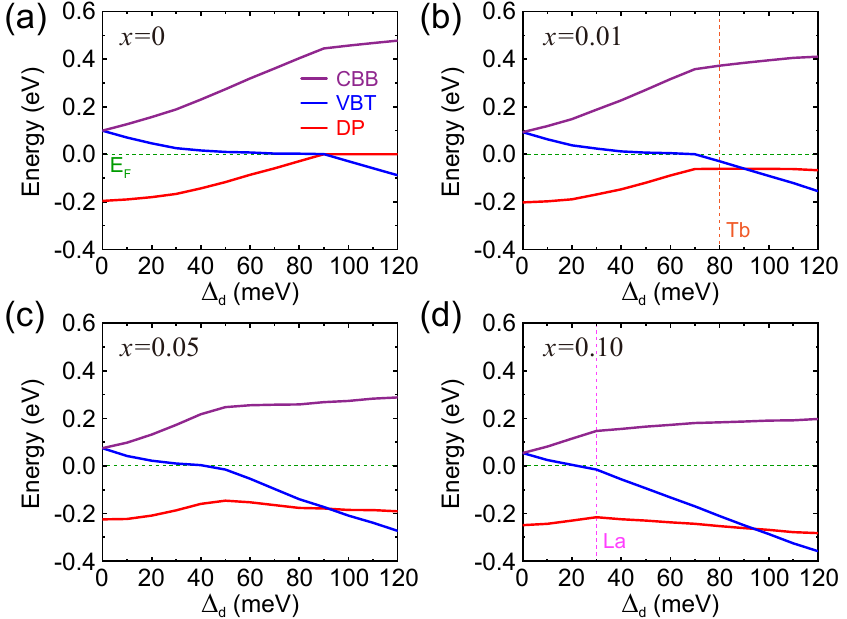} 
\caption{
Order-parameter dependence of the energy positions of the DP, VBT, and CBB in Fig. 1(d) for (a) $x=0$, (b) $x=0.01$, (c) $x=0.05$, and (d) $x=0.10$
where $x$ indicates extra electrons for the filling factor $f=5+x$.
Here, based on cRPA calculations~\cite{liu_electron_2016}, we used $U=1.6, 1.5,$ and $1.4$ eV (and corresponding different $\lambda_{\text{SOC}}^{\text{eff}}$) for $x=0, 0.05,$ and $0.10$, respectively.
Vertical dashed lines marked as Tb and La in (b) and (d) indicate the 3\% Tb-doped and 5\% La-doped Sr$_2$IrO$_4$, respectively. 
}
\end{figure}

As discussed in Refs.~\cite{zhou_correlation_2017} and ~\cite{han_nonsymmorphic_2020}, the optimal parameters used to reproduce ARPES data for the 3\% Tb-doped \cite{han_nonsymmorphic_2020} and 5\% La-doped~\cite{de_la_torre_collapse_2015} Sr$_2$IrO$_4$ are $\Delta_d=80$ and $30$ meV with $x=0.01$ and $x=0.10$, respectively, which are marked as the vertical dashed lines in Fig.~2. 
For the 3\% Tb-doped (5\% La-doped) system, calculated energy positions of the DP, VBT, and CBB are $-61~(-216), -30~(-15),$ and $372~(147)$ meV, respectively.
Given that the energy positions of DP, VBT, and CBB are $-196, 99,$ and $99$ meV for an paramagnetic undoped Sr$_2$IrO$_4$ with $\Delta_d=0$, 
there are significant changes of the energy positions of the three points in the Tb- and La-doped systems, which is expected to result in different electrodynamic properties of Dirac quasiparticles. 
For the sake of comparison, we designate the Dirac line-node state of the paramagnetic undoped Sr$_2$IrO$_4$ ($\Delta_d=0$ and $x=0$) as DLN from now on.

\subsection{Low-energy electrodynamics}

\subsubsection{Intraband optical conductivity}

Using the semiclassical Boltzmann transport theory, we compare the intraband optical conductivities of the DLN, Tb-doped, and La-doped systems.
For the scattering mechanism, we focus on impurity scattering, which allows us to explain the terahertz experiments on the doped Sr$_2$IrO$_4$, as we will see.
When one considers impurity scattering, there are two typical types of impurities depending on the nature of their Coulomb potentials: long-range Coulomb impurities (or charged impurities) distributed randomly in the background and short-range impurities (e.g., lattice defects, vacancies, and dislocations).
The impurity potential for long-range Coulomb impurities is given by $V(\mathbf{q}) = \frac{2\pi e^2}{ \kappa q  } \frac{1}{\varepsilon(q,T)}$ in 2D momentum space where $\kappa$ is an effective background dielectric constant and $\varepsilon(q,T)$ is the dielectric function, whereas the impurity potential for short-range impurities is given by a constant $V(\mathbf{q}) = V_0$.
It is noteworthy that the temperature dependence in the impurity potential happens only for the long-range Coulomb impurities since the dielectric function $\varepsilon(q,T)$ describes the temperature-dependent screening effect. 

\begin{figure}[b]
\includegraphics[width=86mm]{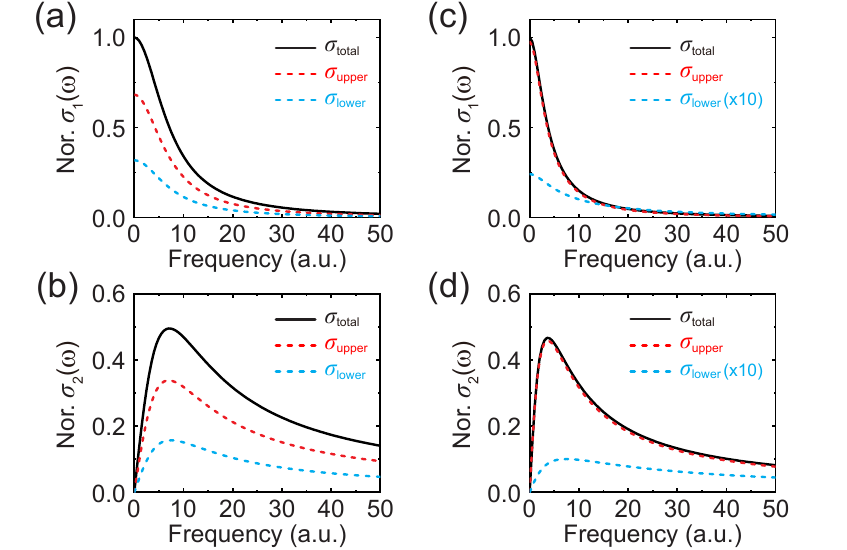} 
\caption{
Calculated real $\sigma_1(\omega)$ and imaginary $\sigma_2(\omega)$ parts of optical conductivity $\sigma_{xx}(\omega)$ at 300 K for the (a),(b) DLN and (c),(d) Tb-doped systems.
Here, in each system, $\sigma_1(\omega)$ and $\sigma_2(\omega)$ are normalized by $\sigma_1(\omega=0)$.
We have considered the two bands near the Fermi level [Figs. 1(a) and 1(b)] that are designated as $\sigma_{\text{upper}}$ and $\sigma_{\text{lower}}$.
Note that $\sigma_{xx}(\omega)=\sigma_{yy}(\omega)$ in our systems, because we have two inequivalent Dirac points at the $X$ and $Y$ points approximately related by $C_{4z}$. 
}
\end{figure}

We first neglect the screening effect [$\varepsilon(q,T)=1$; non-zero screening effect will be treated later] and thus consider only the energy dependence of the scattering rate to calculate the temperature-dependent optical conductivity $\sigma(\omega)$, 
where the temperature dependence of $\sigma(\omega)$ arises from the energy averaging over the Fermi-Dirac distribution function [Eq. (\ref{eq:optical_cond})].
As a result, we deal with two impurity potentials of the bare Coulomb (BC) potential $V_{\text{BC}}(\mathbf{q})= \frac{2\pi e^2}{ \kappa q  } $ and zero-range delta-function (ZD) potential $V_{\text{ZD}}(\mathbf{q})=V_0$.
Note that $V_{\text{BC}}(\mathbf{q})$ is given by substituting $\varepsilon(q,T)=1$ into the impurity potential of long-range Coulomb impurities whereas $V_{\text{ZD}}(\mathbf{q})$ is identical to the impurity potential of short-range impurities. 
We then obtain the energy dependence of scattering rates as $\gamma_{\text{BC}} \propto \epsilon_{\mathbf k}^{-n}$ and $\gamma_{\text{ZD}} \propto \epsilon_{\mathbf k}^{n}$,
where $\epsilon_{\mathbf k}$ is the energy and $n=1/2~(n=1)$ for the DLN system (Tb- and La-doped systems).
Here, the derivation is performed for the anisotropic effective Hamiltonian at the $X$ point constructed from the TB+SOC+$U$+$\Delta$ band structures (see details in Appendices~\ref{Appendix:sec1} and~\ref{Appendix:sec2}).
Note that, the energy ($E$) dependences of $\gamma_{\text{BC}}$ and $\gamma_{\text{ZD}}$ obtained for the Tb- and La-doped systems are the same as those obtained for graphene~\cite{das_sarma_charge_2015}.

Figure~3 shows intraband optical conductivities $\sigma(\omega)$ at 300 K calculated by using the energy-dependent scattering rate $\gamma_{\text{BC}}$ for the DLN and 3\% Tb-doped systems with their TB+SOC+$U$+$\Delta$ band structures. 
Here, we have considered the two bands near the Fermi level [Figs. 1(a) and 1(b)], i.e., $\sigma_{\text{upper}}$ and $\sigma_{\text{lower}}$, which are the only relevant bands in our temperature range.
For the DLN [Figs. 3(a) and 3(b)], $\sigma_{\text{upper}}$ as well as $\sigma_{\text{lower}}$ ($\sigma_{\text{upper}} \sim 2\sigma_{\text{lower}}$) contributes to the total optical conductivity $\sigma_{\text{total}}$, reflecting the dispersing DLN along the BZ boundary. 
On the other hand, for the Tb case [Figs. 3(c) and 3(d)], $\sigma_{\text{lower}}$ is negligibly small and thus $\sigma_{\text{total}} = \sigma_{\text{upper}}$.
Note that, here, considering the energy positions of the DP, VBT, and CBB as well as the Fermi-Dirac distribution function, 
we find that only the linear parts near $X$ and $Y$ points of the upper band [Fig. 1(b)] contribute to $\sigma_{\text{upper}}$.
For the 5\% La-doped system, we also find that $\sigma_{\text{upper}}$ is dominant over $\sigma_{\text{lower}}$, similar to the Tb case.
Despite the difference in details, calculated real $\sigma_1(\omega)$ and imaginary $\sigma_2(\omega)$ parts of the optical conductivity for the three systems show a typical Drude response.
For $\gamma_{\text{ZD}}$, all three systems also show typical Drude responses in $\sigma(\omega)$.

\begin{figure}[b]
\includegraphics[width=86mm]{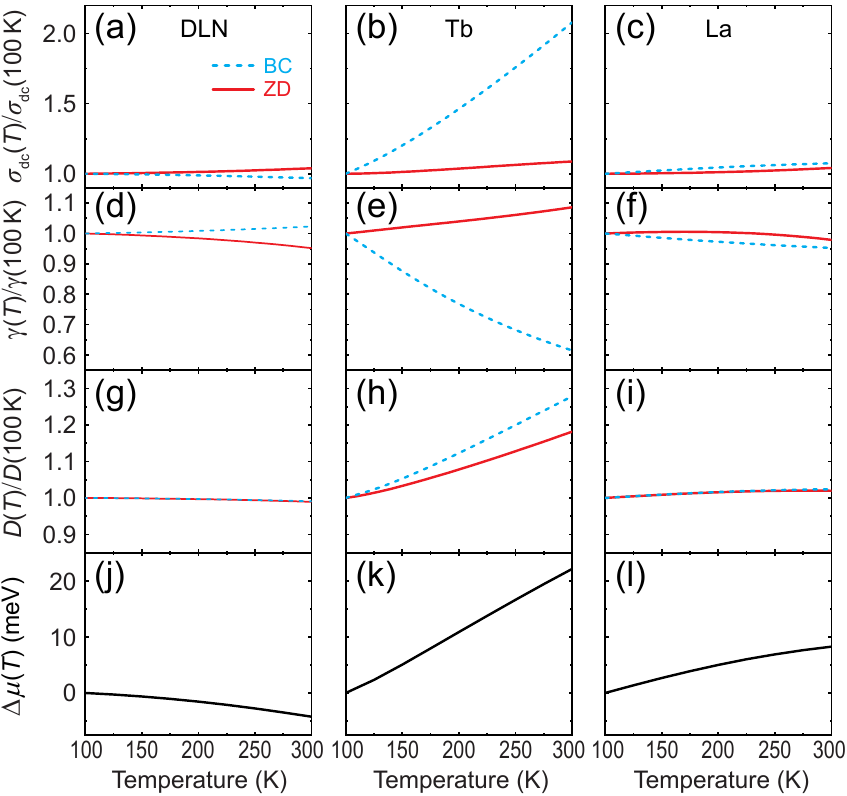} 
\caption{
Calculated temperature-dependent (a)--(c) dc conductivity $\sigma_{\text{dc}}(T)$, (d)--(f) scattering rate $\gamma(T)$, (g)--(i) Drude weight $D(T)$, and (j)--(l) chemical potential $\mu(T)$ for the DLN, 3\% Tb-doped, and 5\% La-doped systems.
}
\end{figure}

By using the Drude model $\sigma(\omega) = D/(\gamma-i\omega) $, we extract the three temperature-dependent electrodynamic parameters of the dc conductivity $\sigma_{\text{dc}}(T)~[=\sigma(\omega=0)]$, scattering rate $\gamma(T)$, and Drude weight $D(T)$ for both the BC and ZD potentials (Fig. 4).
Depending on the impurity potentials (BC or ZD), the DLN and Tb-doped systems show distinct behaviors in the temperature dependence of the electrodynamic parameters, while the La-doped system shows almost the same and featureless tendency irrespective of the BC or ZD potential.
Explicitly, let us discuss $\sigma_{\text{dc}}(T)$ and $\gamma(T)$ first. 
For the DLN, the BC (ZD) potential gives rise to decreasing (increasing) $\sigma_{\text{dc}}(T)$ and increasing (decreasing) $\gamma(T)$ with the temperature $T$, albeit their variations are small.
However, for the Tb-doped system, variations of $\sigma_{\text{dc}}(T)$ and $\gamma(T)$ between the BC and ZD potentials are remarkably different from each other and even definitely show opposite behaviors in $\gamma(T)$. 
Note that, for the Tb-doped case, the BC potential leads to the opposite temperature dependence of the three electrodynamic parameters to those of the DLN. 

For $D(T)$, both the BC and ZD potentials show the same tendency with $T$ for all three systems (even the same magnitudes of $D(T)$ between the BC and ZD potentials in the DLN and La-doped system).
It is consistent with the $T$ dependence of chemical potentials $\mu(T)$ in Figs. 4(j)--(l) where $\mu(T)$ is obtained by the carrier density conservation.
The differences in the $T$ dependence of $\mu (T)$ between Dirac semimetal phases arise from the energy positions of the DP, VBT, and CBB (Fig. 2) the detailed roles of which are discussed below.
The calculated temperature-dependent $\mu(T)$ of the DLN and Dirac point-node (DPN; Tb- and La-doped systems) states show the opposite $T$ dependence.
Specifically, $\mu (T)$ of the DLN decreases with increasing $T$, which is different from a constant $\mu (T)$ of an ideal dispersionless DLN. 
On the other hand, $\mu (T)$ of the DPN is proportional to $T$.
This is opposite to $\mu(T)$ of graphene, basically a DPN, which monotonically decreases as a function of $T$.
The variation of $\mu(T)$ for the Tb-doped system is more than two times larger than that of $\mu(T)$ for the La-doped system, 
contributing to the large variation of $D(T)$ for the Tb-doped system [Figs. 4(h) and 4(i)].

\begin{figure}[b]
\includegraphics[width=86mm]{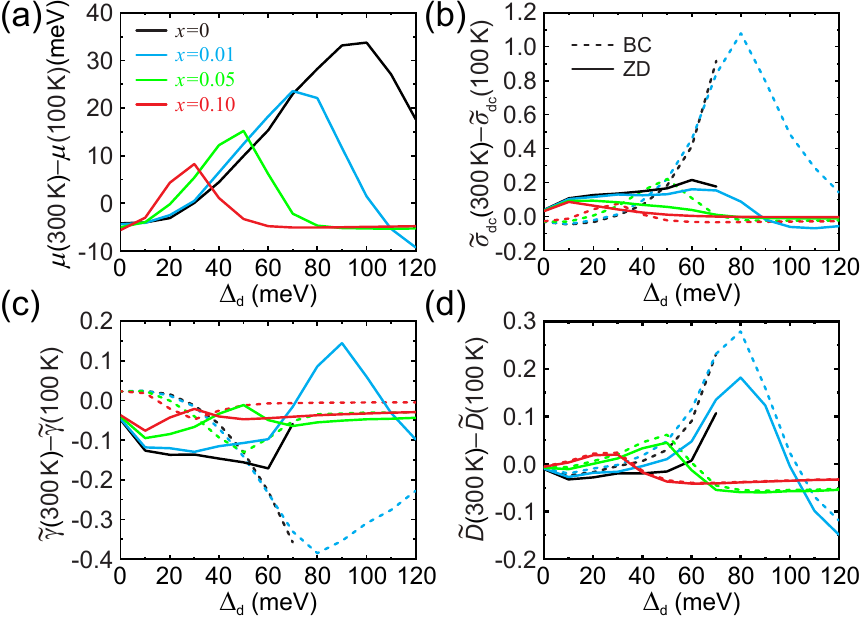} 
\caption{
Order-parameter dependence of the temperature variations of (a) chemical potentials $\mu(T)$, (b) dc conductivity $\tilde{\sigma}_{\text{dc}}(T)\equiv \sigma_{\text{dc}}(T)/\sigma_{\text{dc}}(100~\text{K})$, (c) scattering rate $\tilde{\gamma}(T)\equiv \gamma(T)/\gamma(100~\text{K})$, and
(d) Drude weight $\tilde{D}(T)\equiv D(T)/D(100~\text{K})$ for four different $x=0, 0.01, 0.05, 0.10$.
In (b)-(d), dashed (solid) lines represent data sets calculated by using the BC (ZD) potential.
Note that, for $x=0$, data sets at $\Delta_d \geqq 80$ meV are not drawn because the optical conductivity does not
fit the Drude model due to the close location of the DP to the Fermi level [see Fig. 2(a)].
}  
\end{figure}

To investigate the roles of the DP, VBT, and CBB to the optical properties, we calculate the temperature variations of chemical potentials and electrodynamic parameters as a function of $\Delta_d$ for four different $x$ (Fig. 5).
Since the VBT and CBB give rise to van Hove singularities in the density of states, they dictate the $T$ variation of $\mu(T)$ and thus $T$ variations of electrodynamic parameters $\sigma_{\text{dc}}(T)$, $\gamma(T)$, and $D(T)$ in Fig. 5.
Specifically, for $\mu(T)$ [Fig. 5(a)], it is seen that (i) when the VBT is close to the Fermi level, the $T$ variation is the largest at all $x$ (see also Fig. 2); and (ii) as the CBB goes away from the Fermi level, the $T$ variation becomes large, as seen by comparing the largest values of the $T$ variation between doping levels $x$. 
For $T$ variations of electrodynamic parameters [Figs. 5(b)-5(d)], their peak positions coincide with that of $\mu(T)$, irrespective of the BC or ZD potential, reflecting the roles played by the positions of the VBT and CBB.
Furthermore, they are also affected by the position of the DP: their variations are large when the DP is close to the Fermi level, as seen by comparing results between the doping levels $x$ at the same $T$ variations of $\mu(T)$.

\subsubsection{Comparison with experiments}

Before addressing temperature-dependent screening effects, we compare our results with available terahertz and dc experimental data.
For the 3\% Tb-doped system~\cite{wang_decoupling_2015,han_nonsymmorphic_2020}, it was observed that $\sigma_{\text{dc}}(T)$ and $D(T)$ [$\gamma(T)$] increase (decreases) with $T$ from 120 to 300 K.
This observation agrees surprisingly well with the theoretical results calculated by the BC potential [Figs. 4(b), 4(e), and 4(h)], even with the magnitudes.
We thus conclude that the long-range charged impurities are the dominant scattering mechanism for the 3\% Tb-doped system.
It is noteworthy that this scattering mechanism in the Dirac semimetallic 3\% Tb-doped system can consistently explain both the insulating behavior of temperature-dependent resistivity data~\cite{wang_decoupling_2015} and the metallic Fermi surface observed by ARPES~\cite{han_nonsymmorphic_2020} which were seen to be inconsistent with each other.
Meanwhile, for the 5.5\% La-doped system, decreasing $\sigma_{\text{dc}}(T)$ and nearly constant $\gamma(T)$ and $D(T)$ with $T$ are observed from 200 to 300 K~\cite{han_nonsymmorphic_2020,chen_influence_2015}.
The nearly constant behaviors of $\gamma(T)$ and $D(T)$ are consistent with our theoretical results obtained for both BC and ZD potentials [Figs. 4(f) and 4(i)].
However, for $\sigma_{\text{dc}}(T)$, it is inconsistent with the theoretical result in Fig. 4(c), which might be explained by different scattering mechanisms such as phonon scattering.
Currently, we cannot disclose the main scattering mechanism for the La-doped system due to the lack of experimental data for a wider temperature range,
and thus further study is needed.

\subsubsection{Temperature-dependent screening effect}

We now compare the temperature-dependent screening effects of the Tb- and La-doped systems via the Thomas-Fermi screening wave vector $q_{\text{TF}} (T)$ [see Eq.~(\ref{eq:qTF})]
, where $\varepsilon(q,T) = 1+q_{\text{TF}}(T)/q$.
For this, we consider the low-energy effective Hamiltonian around the $X$ point (see Appendices~\ref{Appendix:subsec1.1} and ~\ref{Appendix:sec2.1.2}).
In the Tb-doped system, $q_{\text{TF}} (T)$ significantly increases from $T= 100$ to $300$ K, indicating a substantial temperature-dependent screening effect [Fig. 6(a)]: The Coulomb potential $V(q)=\frac{2 \pi e^2}{\kappa[q+q_{\text{TF}} (T)]}$ is more screened with increasing $T$.
In contrast, $q_{\text{TF}} (T)$ of the La-doped system in Fig. 6(b) shows small change with $T$, that is, negligible temperature-dependent screening. 
This is due to the deeper position of the Dirac point in the La-doped system compared with the Tb-doped system [see the insets in Figs. 6(a) and 6(b)] as well as smaller changes in $\mu(T)$ for the La-doped system [Figs. 4(k) and 4(l)].

\begin{figure}[b]
\includegraphics[width=86mm]{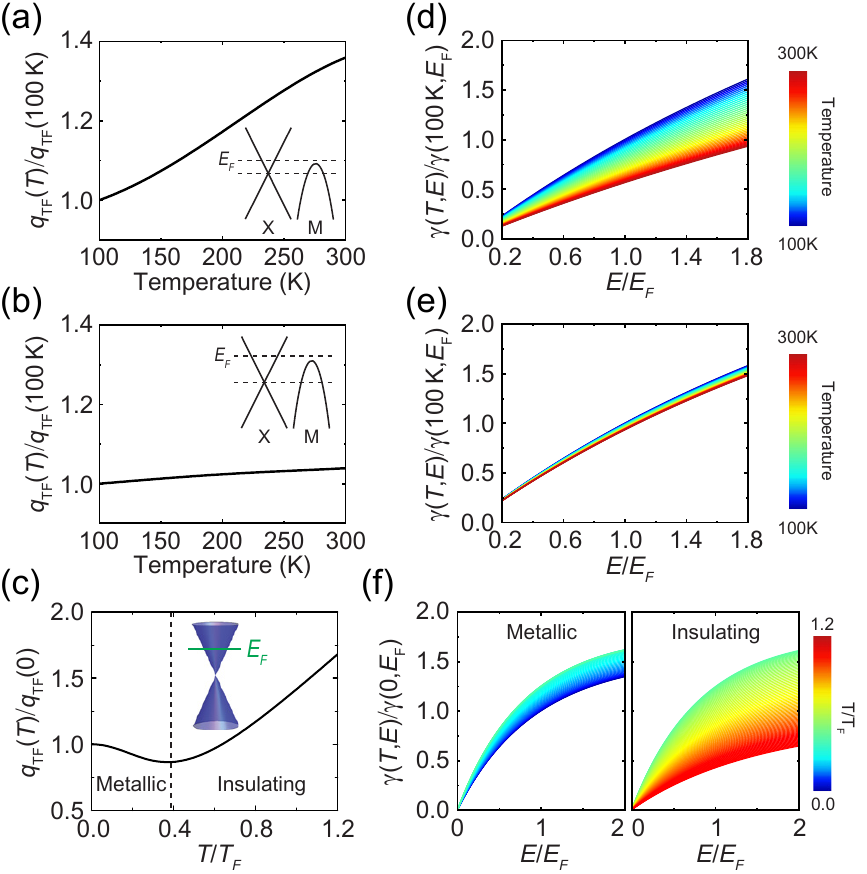} 
\caption{
(a)--(c) Calculated temperature-dependent Thomas-Fermi wave vector $q_{\text{TF}} (T)/q_{\text{TF}} (100K)$ for the (a) Tb-doped, (b) La-doped, and (c) graphene systems. Schematics of the band diagrams with the Fermi level for the three systems are drawn in insets. 
In (c), the dotted line separates metallic and insulating regions.
(d)--(f) Temperature- and energy-dependent scattering rate $\gamma(T,E)$ as a function of $E/E_F$ for the (d) Tb-doped, (e) La-doped, and (f) graphene systems, where the temperature range is indicated in the color bar. 
In (f), we display $\gamma(T,E)$ for two distinct regions. 
Detailed equations and parameters for the three systems are presented in Appendix~\ref{Appendix:sec2.1.2}.
}
\end{figure}

Using $q_{\text{TF}} (T)$, we calculate the anisotropic $T$- and $E$-dependent scattering rate [see Eq.~(\ref{eq:qTF_scattering})];
Figures 6(d) and 6(e) show the angle-averaged $T$- and $E$-dependent scattering rate $\gamma(T,E)$ for the Tb- and La-doped systems, respectively.
At fixed $T$, $\gamma(T,E)$ of both systems increase as a function of $E$, similar to the energy dependence of $\gamma_{\text{ZD}}$ $(\sim \epsilon_{\mathbf k})$.
As $T$ increases, $\gamma(T,E)$ of the Tb-doped system decreases at all $E$, which follows from increasing $q_{\text{TF}} (T)$ as a function of $T$ [Fig. 6(a)].
This is consistent with the obtained scattering rate by the energy averaging of the BC potential [Fig. 4(e)], implying that both the BC potential and temperature-dependent screening effect consistently explain the experimental data in the Tb-doped system.
On the other hand, the variations of $\gamma(T,E)$ with $T$ for the La-doped system are small, as expected by $q_{\text{TF}} (T)$, which is consistent with nearly featureless behaviors of the electrodynamic parameters in Figs. 4(c), 4(f), and 4(i).

To gain more insight, we also plot $q_{\text{TF}} (T)$ and $\gamma(T,E)$ of graphene [Figs. 6(c) and 6(f)], as a reference.
$q_{\text{TF}} (T)$ shows two distinct regimes termed as metallic and insulating: a decreasing region (less screened with increasing $T$) for $T/T_F\lesssim0.4$ and an increasing region (more screened with increasing $T$) for $T/T_F\gtrsim0.4$, where the latter corresponds to the Tb-doped systems.
Such two distinct behaviors lead to metallic and insulating behaviors of the $T$-and $E$-dependent scattering rate $\gamma(T,E)$.
The low-temperature [$T/T_F\sim0.14~(0.43)$ at 100 (300) K] insulating behavior of $\gamma(T,E)$ of the Tb-doped system resembles the high-temperature ($T/T_F\gtrsim0.4$) insulating behavior of $\gamma(T,E)$ of graphene.
Note that the reason why the Tb-doped system does not show metallic behaviors in $\gamma(T,E)$ is the existence of an additional heavy-hole band at the $M$ point [Fig. 1(d)], which is absent in graphene.

\section{summary}

In summary, we have investigated the low-energy electrodynamic properties of various Dirac semimetal phases in the paramagnetic lightly doped Sr$_2$IrO$_4$.
Depending on the $d$-wave order parameter and electron doping level, various Dirac semimetal phases have different energy positions of the Dirac point, valence-band top, and conduction-band bottom and hence show different optical properties.
Using the derived energy dependences of scattering rates originated from short-range and long-range Coulomb impurities, we compare the intraband optical conductivities for the representative Dirac line-node and Dirac point-node phases.
We showed the strong temperature dependence of the electrodynamic parameters in the Tb-doped system while weak dependences for the Dirac line-node and La-doped systems, which are consistent with available dc and terahertz experimental data.
It was also revealed that the screening effect consistently exhibits strong (weak) temperature dependence in the Tb- (La-) doped system.
These differences are mainly attributed to the closer location of the Dirac point to the Fermi level and a large variation of the chemical potential with temperature of the Tb-doped system.
Our work provides the transport and optical fingerprints of each Dirac semimetal phase and signatures of possible phase transition between them, which will stimulate further experimental and theoretical works.

For future work, one interesting research direction is to investigate the phase transition from the Dirac point node to the Dirac line node driven by temperature (or other external stimuli), which is naturally expected considering the temperature dependence of the pseudogap in the lightly doped Mott insulator Sr$_2$IrO$_4$.
Another is to investigate the novel Dirac physics of the correlated Dirac semimetal in doped Sr$_2$IrO$_4$ by tuning the position of the Dirac point via changing doping concentrations or introducing new dopants.
For example, if certain conditions are fulfilled (e.g., pinning of the Dirac point at very near the Fermi level), one might be able to study the Dirac fluid and relativistic hydrodynamics, which was recently observed in graphene~\cite{crossno_observation_2016,lucas_hydrodynamics_2018,gallagher_quantum-critical_2019}, of the correlated Dirac semimetal in doped Sr$_2$IrO$_4$.

\begin{acknowledgments}
This work was supported by NRF through Basic Science Research Programs (NRF-2018R1C1B6007607), the research fund of Hanyang University (HY-2017), and the POSCO Science Fellowship of POSCO TJ Park Foundation.
\end{acknowledgments}

\appendix
\section{Effective Hamiltonian}\label{Appendix:sec1}

In this Appendix, we construct the low-energy effective Hamiltonian in the $J_{\text{eff}}=1/2$ basis.
Using the $k\cdot p$ theory and unitary transformation, we construct Dirac Hamiltonians for the DPN states (Tb- and La-doped systems) near the $X$ and $Y$ points. 
For comparison, we also present the low-energy effective Hamiltonian for the DLN state near the $X$ point.
These effective Hamiltonians are used for scattering rate calculations in Appendix~\ref{Appendix:sec2}.
The effective Hamiltonian of the single-layer Sr$_2$IrO$_4$ in the $J_{\text{eff}}=1/2$ basis is given by \cite{carter_theory_2013,zhou_correlation_2017} 
\begin{align}
\begin{split}
\label{seq:effective H}
H_{\text{eff}} (\mathbf{k}) = \epsilon_1 (\mathbf{k}) 
+ \epsilon_2 (\mathbf{k}) \tau_x 
+  \epsilon_3 (\mathbf{k}) \tau_y \sigma_z 
+ \epsilon_\Delta (\mathbf{k}) \tau_y \sigma_z,
\end{split}
\end{align}
where 
\begin{align}
\begin{split}
\epsilon_1 (\mathbf{k}) & =  
2t_1\left [ \cos ( k_x ) + \cos( k_y ) \right ] + 4  t_{1p}  \cos( k_x   )  \cos ( k_y  ) ,
\\
\epsilon_2 (\mathbf{k}) & =  4 t_2 \cos( k_x /2) \cos( k_y /2),
\\
\epsilon_3 (\mathbf{k}) & =  4 t_3  \cos( k_x/2) \cos( k_y/2),
\\
\epsilon_{\Delta} (\mathbf{k}) & =   -4\Delta_d \sin( k_x/2) \sin( k_y/2).
\end{split}
\end{align}
Here, $t_1, t_{1p}, t_2 $, and $t_3 $ are hopping integrals, 
$\Delta_d$ is the $d$-SODW order parameter, and
$\tau_i$ and $\sigma_i$ $(i = x, y, z)$ are Pauli matrices in the sublattice and $J_{\text{eff}}=1/2$ basis, respectively.
When $\Delta_d=0$ ($\Delta_d \neq 0 $), this effective Hamiltonian $H_{\text{eff}}$ describes the DLN (DPN) state.
Each parameter can be obtained by fitting with the calculated TB+SOC+$U$ band structures.

\subsection{Dirac Hamiltonian for the DPN state}\label{Appendix:subsec1.1}
When the $d$-SODW order is present ($\Delta_d \neq 0 $), there exist Dirac point nodes at the $X$ and $Y$ points as observed in the Tb- and La-doped Sr$_2$IrO$_4$ systems.
To obtain an explicit form of the Dirac Hamiltonian, using the $k\cdot p$ theory, we expand the effective Hamiltonian $H_{\text{eff}}(\mathbf k)$ around $X = (\pi, 0)$.
By the change of variables $k_x \rightarrow  \pi + k_x$ and $k_y \rightarrow   k_y$,
we get
\begin{align}
H_{\text{eff}}  =   -2 t_2 k_x \tau_{x} -2 ( t_3 k_x +\Delta_d  k_y  ) \tau_{y} \sigma_{z}.
\end{align}
Next, we use the following coordinate transformation:
\begin{align}
\begin{split}
k_x' &= k_x \cos \theta  + k_y \sin \theta, 
\\ k_y' &= k_y \cos \theta  -  k_x \sin \theta,
\end{split}
\end{align}
where $\theta =  \frac{1}{2} \tan^{-1}\left(\frac{2 t_3 \Delta_d }{t_2^2 + t_3 ^2 - \Delta_d^2} \right)$.
Then we apply the unitary transformation $H' = U^{\dagger} H U$ with a unitary operator $U = \exp [ - i \phi \tau_z /2]$, where
$\phi  = \tan^{-1} \left ( \frac{ t_3 \cos \theta + \Delta_d \sin \theta}{t_2 \cos \theta } \right )$.
Then we obtain the following anisotropic Dirac Hamiltonian:
\begin{align} \label{seq:DPN effective H}
H' = -  \alpha k_x' \tau_x - \beta k_y' \tau_y \sigma_z,
\end{align}
where 
$\alpha^2  =  2  [ \Delta_d^2 + t_2^2 + t_3^2 + (t_2^2 + t_3^2-\Delta_d^2) \cos 2 \theta +   2 t_3 \Delta_d \sin 2 \theta   ]$ and 
$\beta^2  =  2 [ \Delta_d^2 + t_2^2 + t_3^2 - (t_2^2 + t_3^2-\Delta_d^2) \cos 2 \theta -   2 t_3 \Delta_d \sin 2 \theta ]$.
The energy eigenvalues are given by
\begin{align}
\epsilon(\mathbf k)  = 
\pm  \sqrt{
         \alpha^2 k_x'^2 + \beta^2 k_y'^2
        }.
\end{align}
Similarly, the anisotropic Hamiltonian near the other Dirac point node at $Y = (0, \pi)$ is given by
\begin{align} 
H' = -  \beta k_x' \tau_x - \alpha k_y' \tau_y \sigma_z,
\end{align}
where the eigenvalues are given by
\begin{align}
\epsilon(\mathbf k) = 
\pm  \sqrt{
         \beta^2 k_x'^2 + \alpha^2 k_y'^2
        }
,
\end{align}
where the anisotropic factor is given by $ \eta=\frac{\alpha}{\beta} \approx 1.9 (2.3)$ for the Tb-doped (La-doped) system.
Note that, in this low-energy limit, two anisotropic Dirac cones at $X$ and $Y$ points are related by the $C_{4z}$ rotation.

Because of the mirror symmetry $M_z  =  i \sigma_z \otimes (k_z \rightarrow -k_z)$,
the effective Hamiltonian at the $X$ point for the DPN state in Eq. (\ref{seq:DPN effective H}) can be divided into two sub-Hamiltonians according to the mirror eigenvalues
$\lambda = \pm i $:
\begin{align}
H_{\text{DPN}}^{\pm} = - v_F \hbar \left ( \eta k_x \tau_x \pm   k_y \tau_y  \right ),
\end{align}
where $v_F=\beta / \hbar $.
To make the problem easier, we use the following polar coordinates $(r, \theta)$:
\begin{align}\label{seq:polar_coordinates}
\begin{split}
k_x  & =   \frac{r}{\eta} \cos \theta,
\\
k_y  & =   r \sin \theta.
\end{split}
\end{align}
Then the Hamiltonian is further transformed into a simpler form, i.e., the form of a Dirac Hamiltonian:
\begin{align}\label{eq:Dirac_H_X_pt}
H_{\text{DPN}}^{\pm}  = 
- v_F \hbar r
\left(\begin{array}{c c}
0 & e^{\mp i \theta} \\
e^{ \pm i \theta}  & 0
\end{array}\right)
.
\end{align}
Here, the energy eigenvalue is given by $\epsilon_{s \mathbf k} = s v_F \hbar \sqrt {\eta^2 k_x^2 +  k_y^2} = s v_F \hbar r$,
where $s=+1$ and $-1$ denote the conduction and valence bands, respectively.
The corresponding eigenstate is given by
\begin{align} \label{seq:wave_fun}
\begin{split}
\ket{\phi_{s \mathbf k}^{\pm}} = \frac{1}{ \sqrt2 }
\left(\begin{array}{c}
e^{\mp i \theta } \\  s 
\end{array}\right) .
\end{split}
\end{align}
The velocity
$ v_{\mathbf k}^{(i)} = \frac{1}{\hbar} \frac{ \partial \epsilon_{s \mathbf k}}{\partial k_{i}} $
can be expressed as
\begin{align}
\begin{split}
v^{(x)}_{\mathbf k} &  =   s \eta  v_F \cos \theta,
\\
v^{(y)}_{\mathbf k} &  =   s  v_F \sin \theta.
\end{split}
\end{align}
The Jacobian of the transformation $J(r,\theta)$ is given by
\begin{align}
J(r,\theta) =  
\begin{vmatrix}
 \frac{\partial k_x}{\partial r} &  \frac{\partial k_x}{\partial \theta}
\\
 \frac{\partial k_y}{\partial r} &  \frac{\partial k_y}{\partial \theta}
\end{vmatrix}
= \frac{r}{\eta}.
\end{align}

\subsection{Effective Hamiltonian for the DLN state}

When the $d$-SODW order is absent ($\Delta_d=0$), the paramagnetic Sr$_2$IrO$_4$ corresponds to the DLN semimetal the nodal line of which exists along the whole BZ boundary.
The low-energy effective Hamiltonian up to the quadratic order near the $X$ point is given by
\begin{align}
H_{\text{eff}}  =  D_p + \left(t_1'  -\frac{D_p}{2} \right) k_y^2 + k_x (t_2'\tau_x + t_3'\tau_y\sigma_z).
\end{align}
Here, we ignored the $k_x^2$ term the coefficient of which vanishes in the low-energy region.
Because of the mirror symmetry $M_z  =  i \sigma_z \otimes (k_z \rightarrow -k_z)$,
the above effective Hamiltonian for the DLN state can be divided into two sub-Hamiltonians according to the mirror eigenvalues
$\lambda = \pm i $:
\begin{align}
H_{\text{DLN}}^{\pm}  =  D_p + \left(t_1'  -\frac{D_p}{2} \right) k_y^2 + k_x (t_2'\tau_x \pm t_3'\tau_y).
\end{align}
The energy eigenvalues are given by
\begin{align}\label{eq:H_DLN}
\epsilon(\mathbf k)  = 
D_p + a k_y^2 +s b \abs{k_x},
\end{align}
where $a=t_1' - D_p/2$, $b=\sqrt{t_2'^2+t_3'^2}$, and $s=+1 ~(-1)$ denotes the conduction (valence) band.
The corresponding eigenstate is given by
\begin{align} \label{eq:DLN_eigenstate}
\ket{\phi_{\mathbf k}^{\pm}} = \frac{1}{ \sqrt2 }
\left(\begin{array}{c}
\pm 1 \\  \frac{t_2'+i t_3'}{b}
\end{array}\right) .
\end{align}
Note that the eigenstate is independent of $\mathbf k$ and $s$.
Next, let us consider the following coordinate transformation:
\begin{align}
\begin{split}
& k_x \rightarrow \xi \frac{a}{b}r^2\cos^2\theta,
\\
& k_y \rightarrow r \sin\theta,
\end{split}
\end{align}
where $\xi = +1 ~(-1)$ for $k_x \geqq 0 ~(k_x \leqq 0)$.
Then the energy eigenvalue of the upper band [Eq.~(\ref{eq:H_DLN}); $s=1$] becomes
\begin{align}\label{eq:H_DLN2}
\epsilon(\mathbf k) =  
D_p + a r^2.
\end{align}
The velocity
$ v_{\mathbf k}^{(i)} = \frac{1}{\hbar} \frac{ \partial \epsilon_{ \mathbf k}}{\partial k_{i}} $
can be expressed as
\begin{align}
\begin{split}
v^{(x)}_{\mathbf k} &  =  \xi \frac{b}{\hbar},
\\
v^{(y)}_{\mathbf k} &  =    \frac{2a}{\hbar} r \sin\theta.
\end{split}
\end{align}
The Jacobian of the transformation $J(r,\theta)$ is given by
\begin{align}
J(r,\theta) =  
\begin{vmatrix}
 \frac{\partial k_x}{\partial r} &  \frac{\partial k_x}{\partial \theta}
\\
 \frac{\partial k_y}{\partial r} &  \frac{\partial k_y}{\partial \theta}
\end{vmatrix}
= \frac{2a}{b}r^2\cos\theta.
\end{align}

\section{Scattering rate}\label{Appendix:sec2}

\subsection{Calculation of the scattering rate for the DPN state} 

In order to obtain the optical conductivity $\sigma_{ij}(\omega)$ in Eq. (\ref{eq:optical_cond}), we calculate the scattering rate (or relaxation time).
Within the relaxation-time approximation, the relaxation time in Eqs. (\ref{eq:relaxation_time_integral_iso}) and (\ref{eq:relaxation_anisotropic}) for isotropic and anisotropic systems can be calculated by evaluating $W_{\mathbf{k}\mathbf{k}'}$ in Eq. (\ref{Fermi_Golden});
thus we calculate the matrix element of the impurity potential
$V_{ \mathbf{k}\mathbf{k}' }$:
\begin{align}
\begin{split}
V_{\mathbf{k}\mathbf{k}'}
&= \bra{\mathbf{k}} V  \ket{\mathbf{k}'} 
= \int d\mathbf{r} \braket{\mathbf{k}|\mathbf{r}} V(\mathbf{r}) \braket{\mathbf{r}|\mathbf{k'}}
\\
&= \int d\mathbf{r} e^{i(\mathbf{k'}-\mathbf{k})\cdot\mathbf{r}} V(\mathbf{r}) \braket{\phi_{\mathbf{k}}|\phi_{\mathbf{k'}}}
= V(\mathbf{q})  \braket{\phi_{\mathbf{k}}|\phi_{\mathbf{k'}}}
,
\end{split}
\end{align}
where $\braket{\mathbf{r}|\mathbf{k}} = e^{i\mathbf{k}\cdot\mathbf{r}} \ket{\phi_{\mathbf{k}}}$ in the Bloch basis and $\mathbf{q}=\mathbf{k'}-\mathbf{k}$.
Then we have
\begin{align}
W_{ \mathbf{k}\mathbf{k}' }
= \frac{ 2 \pi  }{ \hbar } n_{\text{imp}}  \abs{ V (\mathbf{q}) }^2  F_{ \mathbf{k}\mathbf{k}' } \delta(\epsilon_{\mathbf{k}} - \epsilon_{\mathbf{k}'}),
\end{align}
where $F_{ \mathbf{k}\mathbf{k}' } =|\braket{\phi_{\mathbf{k}}|\phi_{\mathbf{k'}}}|^2 $ is a square of the overlap function between $\mathbf{k}$ and $\mathbf{k'}$ states in the same band.
Thus, we obtain the following equation of the relaxation time for the anisotropic Dirac Hamiltonian [Eq. (\ref{eq:Dirac_H_X_pt})] with energy $\epsilon_{\mathbf{k}} = v_F \hbar \sqrt {\eta^2 k_x^2 + k_y^2} = v_F \hbar r$ ($s=1$; conduction band):
\begin{align}
\begin{split}
1 = &\frac{ 2 \pi  }{ \hbar } n_{\text{imp}} \int  \frac{dr' d\theta'}{ (2\pi)^2} J(r',\theta') \abs{ V (\mathbf{q}) }^2  F_{ \mathbf{k}\mathbf{k}' } 
\\ 
&\times \delta(v_F \hbar r-v_F \hbar r')
\left(   \tau_{\mathbf k}^{(i)} - \frac{v_{\mathbf k\rq{}}^{(i)}}{v_{\mathbf k}^{(i)}}  \tau_{\mathbf k\rq{}}^{(i)}  \right )
\\
\label{aniso_scattering}
= &\frac{ n_{\text{imp}} }{ 2 \pi   \hbar^2 v_F }  \frac{r}{\eta}  \int   d\theta'  \abs{ V (\mathbf{q}) }^2  \frac{1+\cos(\theta-\theta')}{2} 
\\
&\times \left(   \tau_{\theta}^{(i)} - d_{\theta\theta'}^{(i)}  \tau_{\theta\rq{}}^{(i)}  \right ),
\end{split}
\end{align}
where $F_{ \mathbf{k}\mathbf{k}' }=\frac{1+\cos(\theta-\theta')}{2}$ [see Eq. (\ref{seq:wave_fun})], $d_{\theta\theta'}^{(i)}\equiv\frac{v_{\mathbf k\rq{}}^{(i)}}{v_{\mathbf k}^{(i)}}$, and $\tau_{\theta}^{(i)}$ is the angle-dependent relaxation time along the $i$th direction.
For isotropic systems ($\eta=1$), we can obtain the relaxation time by using $\tau_{\theta}^{(i)} = \tau$.

\subsubsection{Energy-dependent scattering rate}\label{Appendix:sec2.1.a}

Now, we calculate the anisotropic scattering rate by taking into account the two limiting cases of impurity potential:
the BC potential $V_{\text{BC}}(r) \sim \frac{1}{r}$ and the ZD potential $V_{\text{ZD}}(r) \sim \delta(r)$ where $V_{\text{BC}}(\mathbf{q})= \frac{2\pi e^2}{ \kappa q  } $ and $V_{\text{ZD}}(\mathbf{q})=V_0$ in momentum space.

Let us first consider the BC potential. The anisotropic relaxation time in Eq. (\ref{aniso_scattering}) becomes
\begin{align}
\begin{split}
1 = &\frac{ n_{\text{imp}} }{ 2 \pi   \hbar^2 v_F}  \frac{r}{\eta}  \int   d\theta'   \left(   \tau_{\theta}^{(i)} - d_{\theta\theta'}^{(i)}  \tau_{\theta\rq{}}^{(i)}  \right ) \abs{ \frac{2\pi e^2}{ \kappa q  } }^2
\\
&\times \frac{1+\cos(\theta-\theta')}{2} 
\\
= &\frac{  2\pi e^4  n_{\text{imp}}}{  \hbar^2   \kappa^2 v_F} \frac{1}{r}  \int   d\theta'   \left(   \tau_{\theta}^{(i)} - d_{\theta\theta'}^{(i)}  \tau_{\theta\rq{}}^{(i)}  \right )
\\
&\times \frac{1+\cos(\theta-\theta')}{2\eta\left[\frac{1}{\eta^2}(\cos\theta'-\cos\theta)^2+(\sin\theta'-\sin\theta)^2\right]},
\end{split}
\end{align}
where $q^2=r^2 \left[\frac{1}{\eta^2}(\cos\theta'-\cos\theta)^2+(\sin\theta'-\sin\theta)^2\right]$.
By introducing the dimensionless relaxation time, 
\begin{align}\label{dimless_def}
\tilde{\tau}_{\theta}^{(i)} =  \frac{  2\pi e^4 n_{\text{imp}}  }{   \hbar^2  \kappa^2 v_F} \frac{1}{r} \tau_{\theta}^{(i)},
\end{align}
we have
\begin{align}\label{dimless_relax_eq}
1=  \int_{0}^{2\pi}   d\theta' \omega(\theta,\theta') \left(   \tilde{\tau}_{\theta}^{(i)} - d_{\theta\theta'}^{(i)}  \tilde{\tau}_{\theta\rq{}}^{(i)}  \right ),
\end{align}
where $\omega(\theta,\theta')\equiv \frac{1+\cos(\theta-\theta')}{2\eta\left[\frac{1}{\eta^2}(\cos\theta'-\cos\theta)^2+(\sin\theta'-\sin\theta)^2\right]}$.
This dimensionless equation can be numerically solved by discretizing $\theta$ to $\theta_n ~ (n=1,2,\cdots,N)$ with an interval $\Delta\theta = 2\pi/N$.
Along the $x$ direction, where $d_{\theta\theta'}^{x}=\frac{\cos\theta'}{\cos\theta}~ (\cos\theta\neq0)$, Eq. (\ref{dimless_relax_eq}) then becomes
\begin{align}
\begin{split}
\cos\theta_n &=  \sum_{n'} \Delta\theta' \omega(\theta_n,\theta_{n'}) \left(   \tilde{\tau}_{\theta_n}^{x} \cos\theta_n -   \tilde{\tau}_{\theta_{n'}}^{x} \cos\theta_{n'} \right )
\\\label{discretized_linear}
&= \frac{2\pi}{N} \sum_{n'}  \omega_{nn'} \left(   \tilde{\tau}_{n}^{x} \cos\theta_n -   \tilde{\tau}_{n'}^{x} \cos\theta_{n'} \right )
,
\end{split}
\end{align}
where the index of summation $n'$ runs from $1$ to $N$, except for $n' = n$. 
From this, we generate $N$ equations by inserting $n$ from $1$ to $N$.
Then the set of relaxation times $\{\tilde{\tau}_{n}^{x}\}$, and thus the angle-dependent relaxation time $\tau_{\theta}^x$, can be easily obtained by solving $N$-coupled linear equations.
Similarly, one can obtain $\tau_{\theta}^{y}$ where $d_{\theta\theta'}^{y}=\frac{\sin\theta'}{\sin\theta}~ (\sin\theta\neq0)$.
Note that the angle-dependent scattering rate of the BC potential $\gamma_{\text{BC}}^{(i)}(\theta) = 1/\tau_{\theta}^{(i)}$ is inversely proportional to energy $\epsilon_{\mathbf{k}} (=\hbar v_F r)$ as seen in Eq. (\ref{dimless_def}), irrespective of the angle $\theta$.
This is consistent with the following scattering rate of the BC potential obtained for the isotropic system, graphene~\cite{hwang_screening-induced_2009}:
\begin{align}
\frac{1}{\tau_{\text{BC}}^{\text{iso}}}=
\gamma_{\text{BC}}^{\text{iso}}=
\frac{\pi^2  e^4 n_{\text{imp}}}{ \hbar  \kappa^2}\frac{1}{\epsilon_{\mathbf k}}.
\end{align}

Next, we consider the ZD potential.
The anisotropic relaxation time in Eq. (\ref{aniso_scattering}) becomes
\begin{align}
\begin{split}
1 &= \frac{ n_{\text{imp}} }{ 2 \pi  \hbar^2  v_F}  \frac{r}{\eta}  \int   d\theta' \abs{V_0 }^2  \frac{1+\cos(\theta-\theta')}{2} \left(   \tau_{\theta}^{(i)} - d_{\theta\theta'}^{(i)}  \tau_{\theta\rq{}}^{(i)}  \right )
\\
&= \frac{ n_{\text{imp}} V_0^2}{ 2 \pi   \hbar^2 v_F}  r  \int   d\theta' \frac{1+\cos(\theta-\theta')}{2\eta}  \left(   \tau_{\theta}^{(i)} - d_{\theta\theta'}^{(i)}  \tau_{\theta\rq{}}^{(i)}  \right ).
\end{split}
\end{align}
Let us introduce the following dimensionless relaxation time:
\begin{align}
\tilde{\tau}_{\theta}^{(i)} =  \frac{ n_{\text{imp}} V_0^2}{ 2 \pi  \hbar^2  v_F}  r ~\tau_{\theta}^{(i)}.
\end{align}
Then we have
\begin{align}
1=  \int_{0}^{2\pi}   d\theta' \omega(\theta,\theta') \left(   \tilde{\tau}_{\theta}^{(i)} - d_{\theta\theta'}^{(i)}  \tilde{\tau}_{\theta\rq{}}^{(i)}  \right ),
\end{align}
where $\omega(\theta,\theta')\equiv \frac{1+\cos(\theta-\theta')}{2\eta}$.
This equation can also be solved by discretizing $\theta$ to $\theta_n$ as we did for the BC potential case.
Interestingly, we obtain
\begin{align}
\frac{1}{\tau_{\theta}^{(i)}} =
\frac{1}{\tau^{(i)}} =
\gamma_{\text{ZD}}^{(i)}=
 \frac{ n_{\text{imp}} V_0^2}{ 4  \hbar^2 v_F}\frac{1}{\eta} r 
 = \frac{1}{\eta}\gamma_{\text{ZD}}^{\text{iso}},
\end{align}
where $\gamma_{\text{ZD}}^{\text{iso}}$ is the scattering rate of the ZD potential obtained for the isotropic system,
which indicates that $\gamma_{\text{ZD}}^{(i)}$ is independent of the polar angle $\theta$. 
In contrast to the BC potential case, there is no anisotropy in the scattering rate due to the absence of the momentum transfer $\mathbf q$ in $V_{\text{ZD}}(\mathbf{q})$.
Note that the scattering rate $\gamma_{\text{ZD}}^{(i)} $ is proportional to the energy $\epsilon_{\mathbf{k}} $.

\subsubsection{Scattering rate due to temperature-dependent screening}\label{Appendix:sec2.1.2}

Using the effective Hamiltonian at the $X$ point in Eq. (\ref{eq:Dirac_H_X_pt}), we consider the finite-temperature screening effect arising from Dirac fermions around charged impurities within the random-phase approximation (RPA).
In the RPA, the static dielectric function $\varepsilon(q,T)$ is given by $\varepsilon(q,T) = 1 + v_c (q) \Pi(q,T)$ 
where $v_c(q) = 2\pi e^2 / \kappa q$ and $\Pi (q,T)$ is the finite-temperature polarizability function given by the bare bubble diagram~\cite{hwang_screening-induced_2009}
\begin{align}
\Pi(q,T)
=-\frac{g}{A}\sum_{s,s'} \sum_{\mathbf{k}} \frac{f_{s \mathbf k}-f_{s' \mathbf{k'}}}{\epsilon_{s \mathbf k}-\epsilon_{s'\mathbf{k'}}} F_{s\mathbf k,s'\mathbf k'},
\end{align}
where $A$ is the area of the system and $f_{s \mathbf k} = \{\exp[\beta(\epsilon_{s \mathbf k}-\mu(T))]+1\}^{-1}$ with $\beta=1/k_BT$.
Given that $\epsilon_{s\mathbf{k}} =s v_F \hbar  \sqrt {\eta^2 k_x^2 + k_y^2} = s v_F \hbar r$ and $F_{s\mathbf k,s'\mathbf k'} = |\braket{\phi_{s\mathbf{k}}|\phi_{s'\mathbf{k'}}}|^2 = \frac{1+ss' \cos(\theta-\theta')}{2}$ in our DPN state,  
we have 
\begin{align}\label{RPA_pol_aniso}
\Pi(q,T)
=-\frac{g}{\eta}\sum_{s,s'} \int \frac{ r dr d\theta}{(2\pi)^2} \frac{f_{s \mathbf k}-f_{s' \mathbf{k'}}}{\epsilon_{s \mathbf k}-\epsilon_{s' \mathbf{k'}}}  \frac{1+ss' \cos(\theta-\theta')}{2},
\end{align}
where we have changed the summation to an integral in the polar coordinate $(r,\theta)$.
Note that the polarizability function for our anisotropic Dirac Hamiltonian only differs by the $\eta$ factor from the isotropic graphene case~\cite{hwang_screening-induced_2009}.
The normalized polarizability function $\tilde{\Pi}(q,T)=\Pi(q,T)/D_0$, where $D_0\equiv gE_F/2\pi \hbar^2v_F^2 \eta$ is the density of states at the Fermi level, has the same form as graphene.
Therefore, we use the following dimensionless polarizability function $\tilde{\Pi}(q,T)$, derived in graphene~\cite{hwang_screening-induced_2009}:
\begin{align}
\begin{split}
\tilde{\Pi}(q,T)
&=  \frac{\pi}{8}\frac{q}{r_F} + \frac{\mu(T)}{\epsilon_F} + 2 \frac{ T}{T_F} \ln(1+e^{-\beta\mu}) 
\\
&-\frac{1}{r_F}\int_{0}^{q/2}dr 
\frac{\sqrt{1-(2r/q)^2}}{1+\exp[\beta(\epsilon_k-\mu(T))]} 
\\
&-\frac{1}{r_F}\int_{0}^{q/2}dr 
\frac{\sqrt{1-(2r/q)^2}}{1+\exp[\beta(\epsilon_k+\mu(T))]}.
\end{split}
\end{align}

In the $q\rightarrow0$ limit (Thomas-Fermi approximation), 
the dielectric function becomes $\varepsilon(q,T)= 1 + q_{\text{TF}}(T)/q$, where
the Thomas-Fermi wave vector $q_{\text{TF}}(T)$ is given by 
\begin{align}
\begin{split}  \label{eq:qTF}
q_{\text{TF}}(T) &\equiv \lim_{q\rightarrow0} q v_c D_0 \tilde{\Pi}(q,T) 
\\
& = \frac{4 r_F \alpha}{\eta} \left[ \frac{\mu(T)}{\epsilon_F} + 2 \frac{ T}{T_F} \ln(1+e^{-\beta\mu}) \right],
\end{split}
\end{align}
where $\alpha=e^2 / \kappa \hbar v_F$ is the effective fine-structure constant which characterizes the interaction strength of charged impurities.
Then Eq. (\ref{aniso_scattering}) becomes 
\begin{align}
\begin{split} \label{eq:qTF_scattering}
1 = &\frac{ n_{\text{imp}} }{ 2 \pi   \hbar^2 v_F }  \frac{r}{\eta}  \int   d\theta' \left(   \tau_{\theta}^{(i)} - d_{\theta\theta'}^{(i)}  \tau_{\theta\rq{}}^{(i)}  \right ) \abs{ V (\mathbf{q}) }^2  \frac{1+\cos(\theta-\theta')}{2} 
\\
= &\frac{ n_{\text{imp}} }{ 2 \pi   \hbar^2 v_F}  \frac{r}{\eta}  \int   d\theta' \left(   \tau_{\theta}^{(i)} - d_{\theta\theta'}^{(i)}  \tau_{\theta\rq{}}^{(i)}  \right )
\\
&\times \abs{ \frac{2\pi e^2}{ \kappa [q+q_{\text{TF}}(T)]  } }^2 \frac{1+\cos(\theta-\theta')}{2}  
\\
= &\frac{  2\pi e^4 n_{\text{imp}}  }{  \hbar^2   \kappa^2 v_F} \frac{r}{2\eta}  \int   d\theta'  \left(   \tau_{\theta}^{(i)} - d_{\theta\theta'}^{(i)}  \tau_{\theta\rq{}}^{(i)}  \right )
\\
&\times \frac{1+\cos(\theta-\theta')}{\left|r\sqrt{\frac{1}{\eta^2}(\cos\theta'-\cos\theta)^2+(\sin\theta'-\sin\theta)^2}+q_{\text{TF}}(T)\right|^2}
.
\end{split}
\end{align}
Here, by introducing the dimensionless relaxation time and discretizing $\theta$ to $\theta_n$,
we can obtain the anisotropic scattering rate.
Note that there is a temperature dependence in the scattering rate, contrary to the previously discussed energy-dependent scattering rate for the BC and ZD potentials.
The calculated $T$-and $E$-dependent scattering rates, the angle dependence of which is averaged, for the Tb-doped and La-doped systems are presented in Figs. 5(d) and 5(e), respectively.
We set the effective background dielectric constant $\kappa = 1$ (and thus $\alpha=5.9$) for simplicity, and the choice of $\kappa$ does not qualitatively change the temperature dependence of the scattering rates for the Tb- and La-doped systems.
For comparison, we also plot the calculated $T$-and $E$-dependent scattering rate for graphene with the energy $\epsilon_{\mathbf k}=\hbar v_F k$ [Fig. 5(f)], which is given
by~\cite{hwang_screening-induced_2009}
\begin{align}
\begin{split}\label{eq:graphene_scattering}
\gamma(T,E) = \frac{ 4\pi e^4 n_{\text{imp}}}{\hbar \kappa^2} \frac{1}{\epsilon_{\mathbf k}} \int_0^1 dx \frac{x^2 \sqrt{1-x^2}}{[x+\hbar v_F q_{\text{TF}}(T)/2\epsilon_{\mathbf k}]^2},
\end{split}
\end{align}
where $q_{\text{TF}}(T)$ is given by Eq.~(\ref{eq:qTF}) with $\eta=1$. 
Here, we used $\kappa=1~(\alpha = 2.2)$ which corresponds to the suspended graphene in vacuum.

\subsection{Calculation of the scattering rate for the DLN state}

For the DLN state, we only consider the energy-dependent scattering rate.
The integral equation of the relaxation time [Eq. (\ref{eq:relaxation_anisotropic})] for the DLN state with the energy $\epsilon(\mathbf k) = D_p + a k_y^2 + b \abs{k_x} = D_p + a r^2$ [Eqs. (\ref{eq:H_DLN}) and (\ref{eq:H_DLN2}); $s=1$, conduction band] is given by
\begin{align}
\begin{split}
1 =& \frac{ 2 \pi  }{ \hbar } n_{\text{imp}} \int  \frac{dr' d\theta'}{ (2\pi)^2} J(r',\theta') \abs{ V (\mathbf{q}) }^2  F_{ \mathbf{k}\mathbf{k}' } \delta(a r^2 - a r'^2)
\\
&\times \left(   \tau_{\mathbf k}^{(i)} - \frac{v_{\mathbf k\rq{}}^{(i)}}{v_{\mathbf k}^{(i)}}  \tau_{\mathbf k\rq{}}^{(i)}  \right )
\\
=& \frac{ n_{\text{imp}} }{  \pi   \hbar}  \frac{a}{b}  \int dr'  d\theta' r'^2 \cos\theta' \abs{ V (\mathbf{q}) }^2 \frac{1}{2ar}\delta(r'-r) 
\\
&\times \left(   \tau_{\theta}^{(i)} - d_{\theta\theta'}^{(i)}  \tau_{\theta\rq{}}^{(i)}  \right ) 
\\
\label{aniso_scattering_DLN}
=& \frac{ n_{\text{imp}} }{ 2 \pi   \hbar}  \frac{1}{b} r  \int  d\theta'  \cos\theta' \abs{ V (\mathbf{q}) }^2  \left(   \tau_{\theta}^{(i)} - d_{\theta\theta'}^{(i)}  \tau_{\theta\rq{}}^{(i)}  \right ),
\end{split}
\end{align}
where we have used $\delta(a r^2 - a r'^2)=\frac{1}{2ar}[\delta(r'-r)+\delta(r'+r)]$, $F_{ \mathbf{k}\mathbf{k}' }=1$ [see Eq. (\ref{eq:DLN_eigenstate})], and $d_{\theta\theta'}^{(i)}\equiv\frac{v_{\mathbf k\rq{}}^{(i)}}{v_{\mathbf k}^{(i)}}$.

Let us first consider the BC potential. 
The anisotropic relaxation time in Eq. (\ref{aniso_scattering_DLN}) becomes
\begin{align}
\begin{split}
1 =& \frac{ n_{\text{imp}} }{ 2 \pi   \hbar b}  r  \int   d\theta'  \left(   \tau_{\theta}^{(i)} - d_{\theta\theta'}^{(i)}  \tau_{\theta\rq{}}^{(i)}  \right ) \cos\theta' \abs{ \frac{2\pi e^2}{ \kappa q  } }^2  
\\
\label{eq:DLN_BC}
=& \frac{  2\pi  e^4 n_{\text{imp}} }{  \hbar  \kappa^2 b} \frac{1}{r}  \int   d\theta'   \left(   \tau_{\theta}^{(i)} - d_{\theta\theta'}^{(i)}  \tau_{\theta\rq{}}^{(i)}  \right )
\\
&\times \frac{\cos\theta'}{\left[\frac{a^2}{b^2}r^2(\xi'\cos^2\theta'-\xi\cos^2\theta)^2+(\sin\theta'-\sin\theta)^2\right]}
,
\end{split}
\end{align}
where $q^2=r^2\left[\frac{a^2}{b^2}r^2(\xi'\cos^2\theta'-\xi\cos^2\theta)^2+(\sin\theta'-\sin\theta)^2\right]$.
By introducing the dimensionless relaxation time, 
\begin{align}\label{dimless_BC_DLN}
\tilde{\tau}_{\theta}^{(i)} =  \frac{  2\pi e^4  n_{\text{imp}} }{   \hbar \kappa^2 b} \frac{1}{r} \tau_{\theta}^{(i)},
\end{align}
we have
\begin{align}\label{eq:DLN_BC_dimless}
1=  \int_{0}^{2\pi}  d\theta' \omega(\theta,\theta') \left(   \tilde{\tau}_{\theta}^{(i)} - d_{\theta\theta'}^{(i)}  \tilde{\tau}_{\theta\rq{}}^{(i)}  \right ),
\end{align}
where $\omega(\theta,\theta')\equiv \frac{\cos\theta'}{\left[\frac{a^2}{b^2}r^2(\xi'\cos^2\theta'-\xi\cos^2\theta)^2+(\sin\theta'-\sin\theta)^2\right]}$.
This dimensionless equation can be numerically solved by discretizing $\theta$ to $\theta_n ~ (n=1,2,\cdots,N)$ with an interval $\Delta\theta = 2\pi/N$, as we did for the DPN state.
Here, however, we are only interested in the energy dependence of the scattering rate of the BC potential $\gamma_{\text{BC}}$ which is used to calculate the optical conductivity in the main text.
As can be seen in Eq. (\ref{dimless_BC_DLN}), the relaxation time is proportional to $r ~(\sim\sqrt{\epsilon_{\mathbf{k}}})$, and thus $\gamma_{\text{BC}} \propto 1/\sqrt{\epsilon_{\mathbf{k}}}$.
Note that, although there is a term $\frac{a^2}{b^2}r^2(\xi'\cos^2\theta'-\xi\cos^2\theta)^2$ in the argument of the integral in Eq. (\ref{eq:DLN_BC_dimless}), its magnitude is much smaller than the $(\sin\theta'-\sin\theta)^2$ term ($a^2r^2/b^2 < 0.02$ for our range of interest); thus, it is considered that $\frac{a^2}{b^2}r^2(\xi'\cos^2\theta'-\xi\cos^2\theta)^2$ vanishes.

Next, we consider the ZD potential.
The anisotropic relaxation time in Eq. (\ref{aniso_scattering_DLN}) becomes
\begin{align}
1 = \frac{ n_{\text{imp}} V_0^2}{ 2 \pi   \hbar b}  r  \int   d\theta' \cos\theta' \left(   \tau_{\theta}^{(i)} - d_{\theta\theta'}^{(i)}  \tau_{\theta\rq{}}^{(i)}  \right ).
\end{align}
Again, let us introduce the following dimensionless relaxation time:
\begin{align}\label{dimless_ZD_DLN}
\tilde{\tau}_{\theta}^{(i)} = \frac{ n_{\text{imp}} V_0^2}{ 2 \pi   \hbar b}  r~\tau_{\theta}^{(i)}.
\end{align}
Then we have
\begin{align}
1=  \int_{0}^{2\pi}   d\theta' \cos\theta' \left(   \tilde{\tau}_{\theta}^{(i)} - d_{\theta\theta'}^{(i)}  \tilde{\tau}_{\theta\rq{}}^{(i)}  \right ),
\end{align}
which can also be solved by discretizing $\theta$ to $\theta_n$.
We find that there is no anisotropy in the relaxation time, as for the DPN case.
From Eq. (\ref{dimless_ZD_DLN}), one can find that the scattering rate of the ZD potential $\gamma_{\text{ZD}}$ is proportional to $\sqrt{\epsilon_{\mathbf{k}}}$, $\gamma_{\text{ZD}} \propto \sqrt{\epsilon_{\mathbf{k}}}$.

\bibliographystyle{apsrev4-1}
\end{document}